\documentclass[useAMS,usenatbib]{mn2e}
\usepackage{graphicx,multirow,url}
\def\n7{NGC\,7213}
\def\fa{Fe K$\alpha$~}

\def\aj{AJ}%
\def\apj{ApJ}%
\def\apjl{ApJ}%
\def\apjs{ApJS}%
\def\apss{Ap\&SS}%
\def\aap{A\&A}%
\def\mnras{MNRAS}%
\title[\n7:`harder when brighter' X-ray behaviour]{The `harder when brighter' X-ray behaviour of the low luminosity active galactic nucleus \n7}
\author[D.~Emmanoulopoulos et al.]{D.~Emmanoulopoulos,$^{1}$\thanks{E-mail: D.Emmanoulopoulos@soton.ac.uk} I.~E.~Papadakis,$^{2,3}$ I.~M.~M\textsuperscript{c}Hardy,$^{1}$ P.~Ar\'{e}valo,$^{4}$
\newauthor D.~E.~Calvelo,$^{1}$ and P.~Uttley$^{5}$\\
$^{1}$Physics and Astronomy, University of Southampton, SO17 1BJ Southampton, United Kingdom\\
$^{2}$Physics Department, University of Crete, PO Box 2208, 71003 Heraklion, Greece\\
$^{3}$IESL, Foundation for Research and Technology, 71110 Heraklion, Greece\\
$^{4}$Departamento de Ciencias Fisicas, Universidad Andres Bello, Av.Republica 252, Santiago, Chile\\
$^{5}$Sterrenkundig Instituut Anton Pannekoek, University of Amsterdam, Postbus 94249, 1090 GE, Amsterdam, Netherlands}
\begin{document}

\date{Accepted 2012 May 15.  Received 2012 May 14; in original form 2012 April 16}
\pagerange{\pageref{firstpage}--\pageref{lastpage}} \pubyear{2002}
\maketitle

\label{firstpage}
\begin{abstract}
We present the first robust evidence of an anti-correlation between the X-ray photon index, $\Gamma$, and the X-ray luminosity in a single low luminosity active galactic nuclei (LLAGN), \n7. Up to now, such anti-correlation trends have been seen only in large samples of LLAGN that span a wide range of X-ray fluxes, although the opposite behaviour (i.e.\ a positive correlation between  $\Gamma$ and X-ray luminosity) has been extensively studied for individual X-ray bright active galactic nuclei.  For \n7, we use the long-term X-ray monitoring data of \textit{Rossi X-ray Timing Explorer} (\textit{RXTE}), regularly obtained on average every two days from March 2006 to December 2009. Based on our X-ray data, we derive the $\Gamma$ versus flux and the hardness ratio versus flux relations, indicating clearly that \n7 follows a `harder when brighter' spectral behaviour. Additionally, by analysing radio and optical data, and combining data from the literature, we form the most complete spectral energy distribution (SED) of the source across the electromagnetic spectrum yielding a bolometric luminosity of $1.7\times10^{43}$erg s$^{-1}$. Phenomenologically, the SED of \n7 is similar to that of low-ionization nuclear emission-line region. The robust anti-correlation trend that we find between $\Gamma$ and X-ray luminosity together with the low accretion rate of the source, 0.14 per cent that of Eddington limit, make \n7 the first LLAGN exhibiting a similar spectral behaviour with that of black hole X-ray binaries in `hard state'.
\end{abstract}


\section{Introduction}
\label{sect:intro}
\n7 (z=0.005839) is a face on Sa galaxy hosting an active galactic nucleus (AGN). It has been classified as a type I Seyfert by \citet{phillips79}, based on its $H\alpha$ linewidth (full width at zero intensity of $13000$ km s$^{-1}$), but also as a low-ionization nuclear emission-line region (LINER) by \citet{filippenko84},  based on the study of a  variety of optical emission lines which were observed to have a full width at half maximum of  200 to 2000 km s$^{-1}$.\par
\n7 hosts a black-hole with a mass of about $10^8\;{\rm M}_{\sun}$ \citep{woo02}. Its  bolometric luminosity of $L_{\rm bol}=9\times10^{42}$ erg s$^{-1}$\citep{starling05}  suggests a rather low accretion rate of 0.07 per cent of the Eddington luminosity (L$_{\rm Edd}$). As noted by \citet{lobban10}, this value is intermediate between those usually found in type I Seyfert galaxies and LINER's and it is significantly less than the predicted 2 per cent L$_{\rm Edd}$ `critical' rate at which the `soft state' transition appears in black hole X-ray binaries (BHXRBs) \citep{maccarone03}. Finally, based on its radio properties, \n7 belongs to a rare class of extragalactic sources lying between the radio loud and radio quiet AGN (it is one of the 20 sources of the \citet{roy97} sample). In this sense, it can be considered as an extragalactic analogue of the Galactic `hard state' sources.\par
The X-ray spectral behaviour \n7 is also peculiar. \citet{starling05} using the reflection grating spectrometer, on board \textit{XMM-Newton}, found several emission features with no absorption lines; contrary to what is usually observed in type I Seyfert galaxies. The absence of a Compton reflection component from either neutral or ionized material together with the lack of a relativistic \fa line suggests that the inner, optically thick accretion disc in the source may be absent \citep{lobban10}, and replaced by an advection-dominated accretion flow (ADAF) \citep{starling05}. This possibility further supports the `hard state' interpretation of the source.\par
Recently, \citet{bell11} used radio and X-ray observations of \n7, from a long-term observing campaign consisting of several years, to search for correlated X-ray and radio variations that might originate in the so-called `fundamental plane of black hole accretion' \citep{merloni03,koerding06}. \citet{bell11} showed that the average radio and X-ray luminosities fitted well with the global fundamental plane, consistent with a `hard state' identification of this AGN. However, the X-ray versus radio correlation within the monitoring period was weak, with significant intrinsic scatter away from the plane.\par
Numerous studies in the past have shown that the X-ray photon index, $\Gamma$, correlates with the accretion rate, defined as $\xi=L_{\rm X}/L_{\rm Edd}$ (where $L_{\rm X}$ is usually the 2--10 keV X-ray luminosity) in both AGN and BHXRBs. For example, \citet{shemmer06} showed that $\Gamma$ and $\xi$ are positively correlated, using data for a sample of thirty quasars, while \citet{sobolewska09} found that the same positive correlation holds when one studies the spectral variations of individual, luminous Seyfert galaxies. However, the positive $\Gamma-\xi$ correlation may not hold in less luminous AGN. In fact \citet{gu09} found an anti-correlation between $\Gamma$ and $\xi$ in a sample of 55 low-luminosity AGN (LLAGN), and \citet{younes11} reached a similar conclusion for a sample of 13 optically selected LINER's.\par
\citet{wu08} performed a detailed spectral study for six BHXRBs and found that $\Gamma$ anti-correlates with $\xi$ below a `critical' value of $\log(\xi_{\rm crit})=-2.1\pm0.2$. At higher accretion rates, $\Gamma$ and $\xi$ are positively correlated. Similar conclusions were also reached recently by \citet{sobolewska11}. The $\Gamma$ -- $\xi$ relation in AGN may also change from negative to positive at a similar `critical' value, as it was suggested by \citet{wu08}, and \citet{constantin09}. This analogous behaviour between AGN and BHXRBs reinforces the interpretation that LLAGN are analogues of the `hard state' BHXRBs while luminous Seyferts and quasars are the equivalent of the `soft state' BHXRBs. On a first look, during the `hard state', BHXRBs appear to have a constant hardness-ratio, in the commonly used hardness-intensity diagrams \citep[e.g.][]{belloni05}. Nevertheless, detailed analysis of the lowest luminosity regime of this state, known also as `low-hard' state, exhibits clearly an increase of the hardness-ratio for increased fluxes, corresponding to an anti-correlation between $\Gamma$ and $\xi$ \citep{heil12}.\par
The negative and the positive correlations between $\Gamma$ and $\xi$, occurring below and above $\xi_{\rm crit}$, respectively, may be indicative for some `switch' in the emission mechanism as the source's accretion rate increases above $\xi_{\rm crit}$. In fact, it has been suggested \citep{wu08,constantin09,younes11} that this transition could indicate the passage from an ADAF \citep{narayan94,esin97} to standard disc (but see \citet{sobolewska11} for alternative explanations as well).\par
The positive correlation relation between $\Gamma$ versus $\xi$ in AGN has been firmly established statistically by either considering samples of numerous sources \citep[e.g.][]{shemmer06} or large data sets for a few individual sources \citep[e.g.][]{mchardy99,larmer03,sobolewska09}. However, up to now the $\Gamma$ -- $\xi$ anti-correlation has been established using only short X-ray observations of numerous different sources \citep{gu09,constantin09,younes11}, while it has never been observed in a single source.\par
In this work, we use long-term, monitoring \textit{RXTE} data, and present, for the first time, conclusive evidence for such an anti-correlation for an individual LLAGN, namely \n7. We also use archival and proprietary data to construct its average spectral energy distribution (SED), and compare it with the average SED of luminous AGN and LINER's. In Section \ref{sect:obs_reduc} we refer to the observations and data reduction procedures and in Section \ref{ssect:rxte_lc} we present our results consisting of the $\Gamma$ versus flux relation and the spectral energy distribution of the source (SED) of the source. Finally, a discussion together with a summary can be found in Section \ref{sect:discussion}. The cosmological parameters used throughout this paper are: $\rm{H}_0=70$ km s$^{-1}$ Mpc$^{-1}$, $\Omega_{\Lambda}$=0.73 and $\Omega_{\rm m}$=0.27, yielding a luminosity distance to \n7 of 22.12 Mpc (for a corrected redshift, $z_{\rm corr.3K}=0.005145$ to the reference frame defined by the 3 K cosmic microwave background radiation). This value of the luminosity distance appears to be fully consistent
with the redshift-independent Tully-Fisher distance \citep{tully88}.
 
\section{OBSERVATIONS AND DATA-REDUCTION}
\label{sect:obs_reduc}
\subsection{X-ray data}
\n7 has been observed regularly by the \textit{Rossi X-ray Timing Explorer} (\textit{RXTE}) (proposal numbers: 92119, 93139, 94139 and 94342) from 2006 March 3, 02:57:41 (UTC) to 2009 December 30, 00:18:37 (UTC) (on-time: 800506 s). This is the complete set of X-ray observations of \n7, obtained by RXTE, extending the X-ray data of \citet{bell11} by an additional of 100 days. In order to form the most homogeneous long-term data set, we use data obtained only by the proportional counter array 2 (PCU\,2), one of the five Xenon proportional counters forming the proportional counter array \citep{jahoda96}. PCU\,0 and 1 have lost their propane layer, resulting an increased background rate as well as a different detector gain, and PCU\,3 and 4 are usually switched off due to discharge problems.\par
The {\tt Standard-2} (Std2: data with accumulation rate of 16 sec) PCU\,2 data (822 files in total) are extracted and processed using {\tt ftools} \citep{blackburn95} included in {\tt HEAsoft} (ver.~6.11.1), following the standard reduction procedures, provided by the \textit{RXTE} Guest Observer Facility (GOF). For each observation we create a filter file, assembling the scientifically important parameters, using the script {\sc xtefilt} and then by using the {\tt ftool} {\sc fmerge} we merge them all in a single `master' filter file. Then, based on this file, we extract the useful observing time periods, known as good time intervals (GTIs), using the {\tt ftool} {\sc maketime} having the following observational constraints: elevation angle greater than 10$\degr$, a pointing offset of less than 0.02$\degr$, electron contamination less than 0.1, time since south Atlantic anomaly between zero and thirty minutes and time since PCU breakdowns less than -150 s or greater than 600 s. Finally, for each Std2 observation a synthetic background model file is created with the {\tt ftool} {\sc pcabackest} using the `faint background model'.\par
For the production of the background-subtracted light curves, in the 2--4, 5--10 and 2--10 keV energy bands, we use the {\tt ftool} {\sc saextrct} to produce the `source-plus-background' and `background' light curves for each energy band, respectively. We select events obtained during the GTIs registered only from the top xenon layer (X1L, X1R) of PCU\,2, optimizing in this way the signal to noise ratio. Then the corrected background-subtracted light curves are produced by using the {\tt ftool} {\sc lcmath} in units of count-rate. In order to produce a total long-term X-ray light curve in the 2--10 keV energy range, in units of flux i.e.\ erg s$^{-1}$ cm$^{-2}$, we extract for each Std2 observation, and the corresponding synthetic background file, a spectrum together with the PCU\,2's response using again the {\tt ftool} {\sc saextrct} and the script {\sc pcarsp}, respectively. Again, we select events obtained during the GTIs which are registered only from the top xenon layer of PCU\,2. The resulting spectra are then grouped so that each spectral bin contains at least 30 counts, using the {\tt ftool} task {\sc grppha}.\par 
Using the X-ray fitting package {\tt XSPEC} (ver.~12.7.0) \citep{arnaud96} we fit to each X-ray spectrum a power law model assuming photoelectric absorption ({\tt XSPEC} model {\tt wabs}), of a fixed interstellar column-density of $N_{\rm H}=1.1\times 10^{20}$ cm$^{-2}$ \citep[estimated using the {\sc ftool} \textit{nH}, after][]{kalberla05} . Finally, after fixing the fitting parameters (i.e.\ photon index and normalisation) to their the best-fitting values, we obtain the flux values, in the 2--10 keV energy range using the convolution model {\tt cflux}. The errors in the spectral indices and fluxes indicate their 68.3 per cent confidence range, corresponding to a $\Delta \chi^2$ of 1, unless otherwise stated.

\subsection{Spectral energy distribution data}
\label{ssect:sed_data}
In order to construct the SED of \n7, we analyse long term monitoring radio and optical data. Below we describe the data analysis procedures for these observations. We also use the published near-infrared, nuclear fluxes of \citet{hoenig10,asmus11} at 12.27 and 11.25 and 10.49 $\mu$m (these fluxes are listed in Table \ref{tab:fluxEntries} with an asterisk). Finally, we also considered the results from a $130$ ksec \textit{XMM-Newton} observation obtained during 55146--55148 MJD (obs: ID-0605800301, Emmanoulopoulos et al.\,in prep.), the 4.5 years average \textit{Swift}-BAT spectrum \citep[as reported in entry 1185 of table 2 in][]{cusumano10}, and the recent 0.1--100 GeV \textit{Fermi-}LAT upper limit \citep{lobban10}.

\begin{table}
\caption{The radio, near-infrared and optical mean flux values of \n7.}
\label{tab:fluxEntries}
\centering\begin{tabular}{@{}cccc}
\hline
 Energy band & Flux \\
 & mJy \\
\hline
 1.344 GHz  & $121.3 \pm 2.2$  \\
 1.384 GHz & $112.4\pm 9.3$    \\
 2.386 GHz  & $114.9 \pm 3.6$    \\
 2.496 GHz  & $98.1 \pm 4.1$    \\
 4.8 GHz & $135.8 \pm 8.3$       \\
 8.64 GHz  & $150.6 \pm 24.3$     \\
 17 GHz  & $140.4 \pm 3.9$     \\
 19 GHz  & $128.5 \pm 4.1$      \\
 12.27 $\mu$m\textsuperscript{\textasteriskcentered} & $235.8 \pm 128.4$ \\ 
 11.25 $\mu$m\textsuperscript{\textasteriskcentered} & $232.9 \pm 144.5$ \\
 10.49 $\mu$m\textsuperscript{\textasteriskcentered} & $239.1 \pm 22$ \\
 5500 \AA &  $0.49 \pm 0.18$ \\
 4400 \AA & $0.73 \pm 0.15$  \\
\hline
\end{tabular}
\medskip \\
\textsuperscript{\textasteriskcentered} the near-IR measurements taken from \citet{hoenig10,asmus11} as explained in the text.
\end{table}

\subsubsection{Radio data}
We searched the Australian Telescope Online Archive (ATOA) for \textit{Australian Telescope Compact Array} (\textit{ATCA}) observations of \n7's location, selecting a range that covered as many frequencies as possible, opting for the longest duration observation(s) in each case (sometimes yielding multiple useful files, as in the case of 5--10 GHz). While attempts were made to process all designated files, the highest available frequencies, greater than 20 GHz, did not yield useful images. The final data set used in this work was comprised of 21 ATCA observations, from four projects; C782, C1803, C1532 and C1392. \par
The primary calibrator for all these data sets was PKS\,1939-6342 (PKS\,B1934-638). C782 includes observations from 1999 March 03 and frequencies of 1344 MHz (0.3 h on source, average beam size (ABS): $64.7\arcsec\times7.0\arcsec$), 1384 MHz (8.5 h on source, ABS: $21.6\arcsec\times7.2\arcsec$) and 2496 MHz (8.8 h on source, ABS: $7.6\arcsec\times4.9\arcsec$ ), with the antennae in configuration 6C (minimum baseline length of 153 m, maximum of 6 km). The secondary calibrator used was PJS\,J2214-3835 (PKS\,B2211-388A). C1803 included observations from 2008 April 21 and 23 at 1384 MHz and 2386 MHz, with 0.15 and 0.1 h on source respectively. The antenna configuration was 6A (minimum baseline length of 337 m, maximum of 5939 m) and the secondary calibrator used was PKS\,J2235-4835 (PKS\,B2232-488). C1532 contributed all our 4800 MHz (ABS: $17.3\arcsec\times1.8\arcsec$) and 8640 MHz (ABS: $9.64\arcsec\times1\arcsec$) data, with observations covering the dates of 2008 January 05, 24, April 12, November 04 and 20 with on source durations of 1.48, 1.7, 1.5, 2.29 and 1.49 h, respectively. The secondary calibrator used in all cases was PKS\,J2218-5038 (PKS\,B2215-508) and antennae were always in configuration 6A. Finally, C1392 provided our highest frequency ATCA data, at 17 GHz (ABS: $35.5\arcsec\times27.55\arcsec$) and 19 GHz (ABS: $29.15\arcsec\times22.8\arcsec$). Two observations are used, from 2009 September 29 and 30 with 0.02 and 0.03 h on source respectively. The secondary calibrator was PKS\,J2248-3235 (PKS\,B2245-328) and the antenna configuration was H75 (minimum baseline of 31m, maximum of 4408m).\par 
Once imaging was complete, point-source fitting is used to measure flux density at \n7's position $\rmn{RA}(1950)=22^{\rmn{h}}~09^{\rmn{m}}~16\fs26$, $\rmn{Dec.}~(1950)=-47\degr~09\arcmin~59\farcs 95$. All data and image processing is carried out in the radio interferometry data reduction package {\sc MIRIAD} \citep{sault95}.

\subsubsection{Optical data}
We used the \textit{ANDICAM} instrument mounted on the 1.3 m telescope \textit{SMARTS} in Chile to observe \n7 in optical B (4400 \AA) and V (5500 \AA) bands. Observations were performed between August 2006 and August 2011, every 4 days and during each observations two 60 s exposure frames in the B band and two 30 s exposures in the V band were obtained when the target was visible. After discarding the `bad seeing' nights, a total of 266 epochs are accumulated in each band. The field of view of \textit{ANDICAM} is $6\arcmin\times 6\arcmin$ covering both the entire host galaxy and a few bright stars, and the pixel size is $0.37\arcsec$ per pixel. The typical seeing during the campaign was about $1.5\arcsec$. The data reduction for both the B and V band images is the same.\par 
Since \n7 host galaxy is a face-on spiral with a bright nucleus compared to the AGN, galaxy subtraction has to be performed to determine the pure AGN flux. To model correctly the large scale structure we use the data analysis algorithm {\tt GALFIT} (ver.~3.0) \citep{peng10}, fitting a Nuker function. After the fitting, a bright ring $10\arcsec$ from the nucleus is left together with some spiral structure inward of the ring and a resolved core. The core is subtracted using a Moffat function leaving a residual nuclear peak consistent with the stellar point-spread function (PSF), which we identified with the AGN. Aperture photometry is performed with the task {\sc daophot}, of the {\tt IRAF} (ver.~2.15) software system \citep{tody93}, on the stars of the original image and on the modelled nuclear PSF to determine its relative brightness. In order to estimate the flux, the reference stars in turn are calibrated taking their average aperture magnitudes over all reported photometric nights when observations were carried out, corrected for airmass extinction. The average nuclear fluxes that are obtained in this way are $5\times10^{-12}$ erg s$^{-1}$ cm$^{-2}$ at 4400 \AA\ and $3\times10^{-12}$ erg s$^{-1}$ cm$^{-2}$ at 5500 \AA, while the fractional root mean square variability amplitude (including observational noise and intrinsic variability) is around 1 erg s$^{-1}$ cm$^{-2}$ for both bands, respectively.

\section{RESULTS FOR \n7}
\begin{figure}
\includegraphics[width=3.5in]{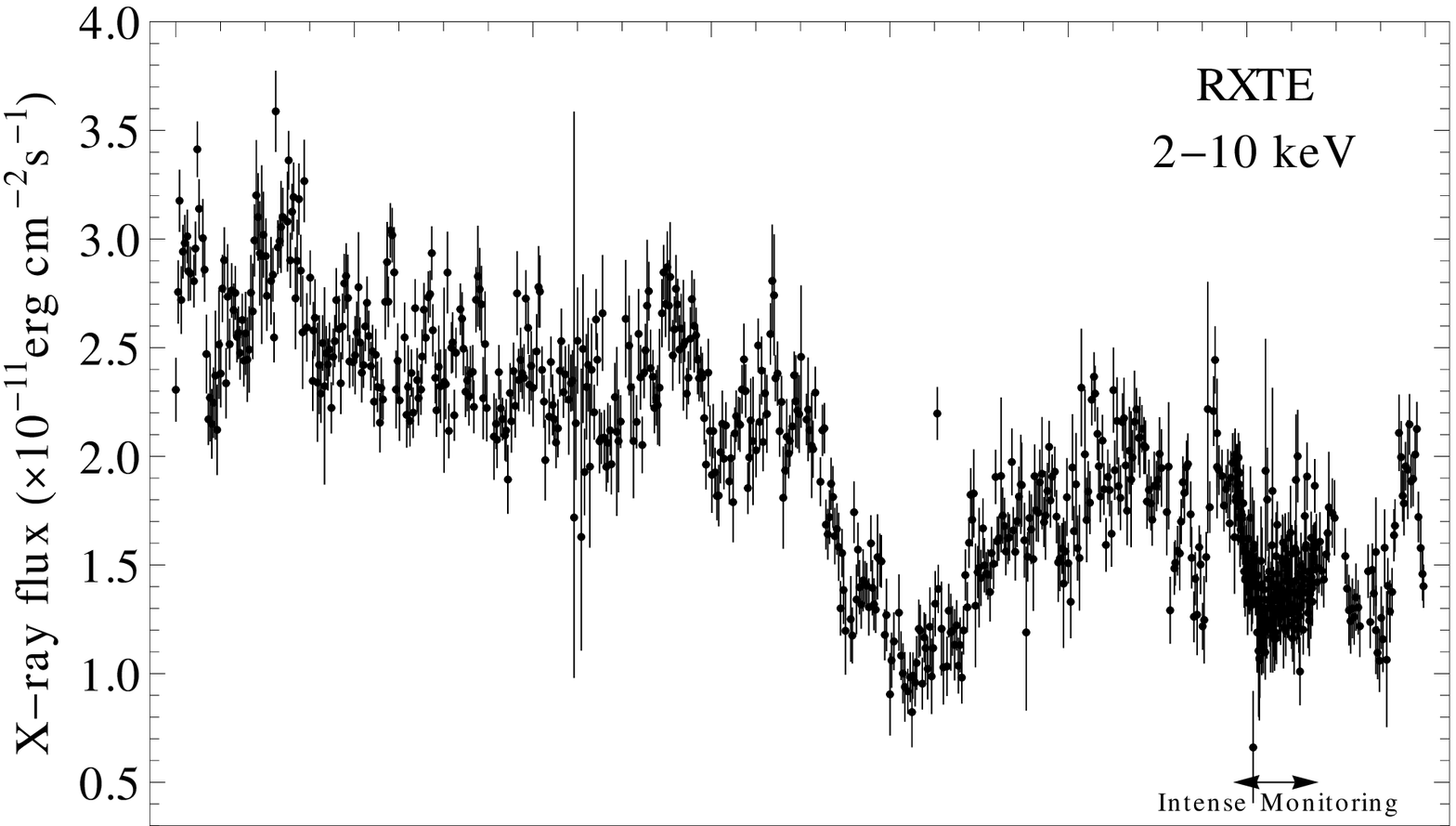}\\[-1.57em]
\hspace*{0.1em}\includegraphics[width=3.555in]{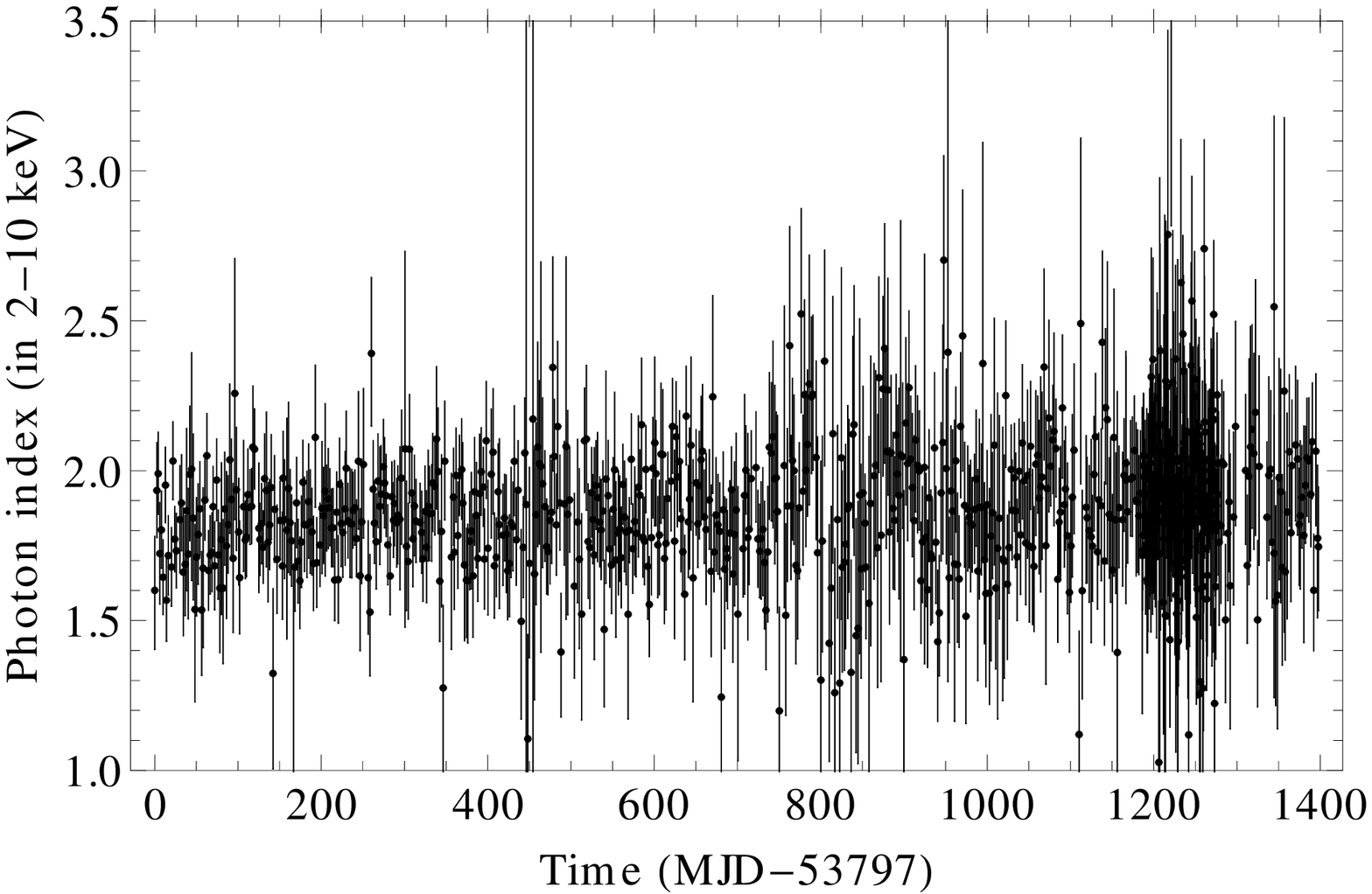}
\caption{The RXTE results of \n7. (Top panel) The long term light curve in the 2--10 keV energy range. The double-headed arrow indicates a period intense monitoring between 54982--55077 MJD. (Lower panel) The photon index of each pointing observation as a function of time.}
\label{fig:rxte_lc_pi}
\end{figure}
\subsection{The X-ray flux -- photon index relation in \n7}
\label{ssect:rxte_lc}
The upper panel of Fig.~\ref{fig:rxte_lc_pi} shows the 2--10 keV, X-ray light curve of \n7. One can observe variations on time scales of days and weeks, superimposed on a flux decreasing trend, from the start to the end of the monitoring campaign. Each measurement on this plot corresponds to one \textit{RXTE} pointing, having a mean duration of $1.06\pm0.25$ ks (the error estimate corresponds to the standard deviation of the exposure times of individual observations). On average \textit{RXTE} observed the source every $\sim 2.3$ days, except from the period between 54982--55077 MJD  (indicated by the double-headed arrow in the upper panel of Fig.~\ref{fig:rxte_lc_pi}), where observations were performed around every half a day. Observations with large uncertainties e.g.\ around (53797+450) MJD, correspond to exposure times less than 0.9 ks.\par
\begin{figure}
\includegraphics[width=3.5in]{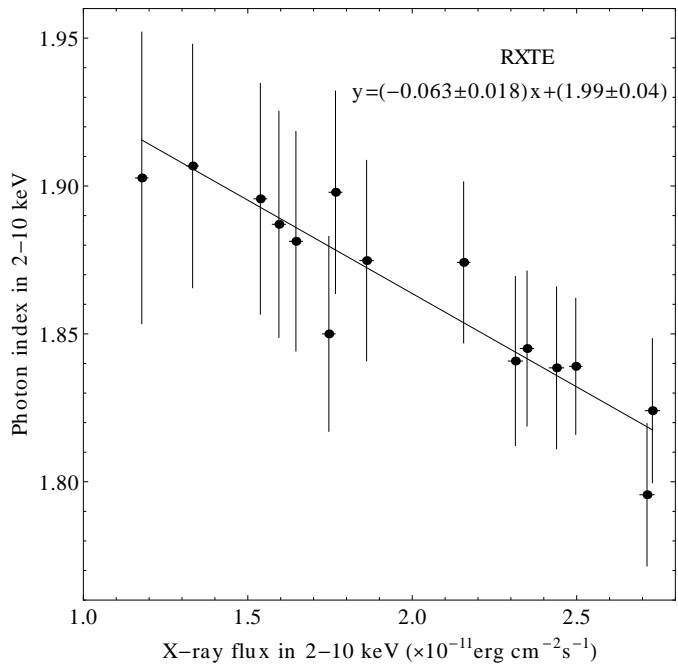}\\[-1.6em]
\caption{The average photon index versus the average X-ray flux of \n7 in bins of 45 consecutive observations. The data suggest a `harder when brighter' relation. The solid line indicates the best-fit linear model to the data taking into account the uncertainties in both coordinates.}
\label{fig:rxte_hard_bright}
\end{figure}

 
The lower panel of Fig.~\ref{fig:rxte_lc_pi} shows the evolution of the $\Gamma$ as a function of time. In order to investigate any possible relationship between flux and $\Gamma$, we group both data sets in bins of 45 consecutive observations (each bin is then $\sim100$ days long). For the intense monitoring period (1185--1280 MJD-53797) we apply the same sampling scheme by selecting only the observations that are separated around 2.3 days and then bin them in bins of 45 consecutive observations. In this way we ensure that the resulting X-ray flux and photon index range are represented by equal number of observations, avoiding the possibility that the larger number of observations in the intense monitoring period could drive the resulting flux/photon index relation. 
We estimate a weighted mean and a weighted error \citep[e.g.][]{bevington92} for both flux and $\Gamma$ in each bin. Fig.~\ref{fig:rxte_hard_bright} shows the resulting average  X--ray flux and $\Gamma$ values, plotted against each other, unveiling a clear `harder when brighter' behaviour.\par

\begin{figure}
\includegraphics[width=3.5in]{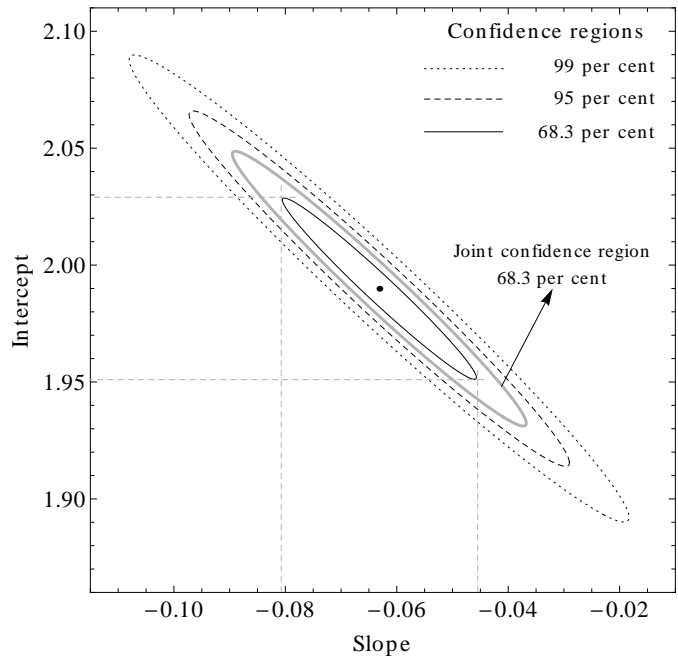}
\caption{The confidence contours for the best fitting model of Fig.~\ref{fig:rxte_hard_bright} taking into account the uncertainties in both coordinates. The black thin lines correspond to the 99, 95 and 68.3 confidence contours and represent the ellipsoids having from the minimum a $\Delta\chi^2$ of 6.63, 3.84 and 1, respectively, for 1 d.o.f. The grey thick line corresponds to the the 68.3 per cent joint confidence contour for the slope and the intercept having from the minimum a $\Delta\chi^2$ of 2.30, for 2 d.o.f. The horizontal and vertical dashed grey lines correspond to the tangents of the 68 per cent confidence region yielding the 68 per cent uncertainty ranges of the slope and intercept respectively.}
\label{fig:rxte_linRegrConf}
\end{figure}

In order to quantify this anti-correlation we first compute the Kendall's $\tau$ rank correlation coefficient \citep{press92a}, using the data plotted in Fig.~\ref{fig:rxte_hard_bright}, yielding a $\tau=-0.81$ with a null hypothesis probability of $2.6\times 10^{-5}$. This results shows that the anti-correlation between spectral slope and flux is highly significant. Then, we fit to the data a linear model, $(y=\alpha x+\beta)$, considering the uncertainties in both coordinates \citep{press92a}. The best-fitting model yields a slope of $\alpha=-0.063\pm0.018$ and an intercept of $\beta=1.99\pm0.04$ with a $\chi^2$ merit function of 2.90 for 13 degrees of freedom (d.o.f.) having a null hypothesis probability of $1.7\times10^{-3}$. Fig.~\ref{fig:rxte_linRegrConf} shows the confidence contour (solid black line) used to estimate the one-standard deviation uncertainties on the best-fitting parameters corresponding to an ellipsoidal region with a $\chi^2$ of 3.90 ($\Delta\chi^2=1$ from the minimum for 1 d.o.f.). In the same plot, we show also the contours for confidences of 95 and 99 per cent as well as the 68.3 per cent joint confidence contour for both the slope and the intercept with the latter corresponding to a $\Delta\chi^2=2.30$ from the minimum for 2 d.o.f. A simple linear regression model, taking into account only the photon index uncertainties, yields equivalent results: $\alpha=-0.063\pm0.008$ and $\beta=1.99\pm0.02$ ($\chi^2$=2.90 for 13 d.o.f). The resulting $\chi^2$, from both methods, are relatively small, indicating that the estimated errors on the average $\Gamma$ within each bin maybe slightly overestimated. We therefore repeat the fit and this time we fit to the data a linear model following the `ordinary least-squares regression of Y on X' routine of \citet{isobe90} which does not take into account the error on the data. The best-fit results from these routines are consistent, within the quoted errors, with the previous results yielding:  $\alpha=-0.072\pm 0.008$ and $\beta=2.00\pm 0.02$.\par
We have performed several sanity checks in order to test the sensitivity of our results to the binning scheme. We have considered alternative binning of 20,60 and 100 consecutive observations as well as considering all the data from the intense monitoring period. Also we have performed data-resampling using jackknifing \citep{shao95} by selecting randomly sub-samples from the data set and binning them in the flux/photon index plane. All the results are extremely consistent with each other. Finally, in order to ensure that the overall anti-correlation trend is not induced by any sort of statistical dependence between the flux and $\Gamma$ \citep[e.g.][]{vaughan01}, we measured the eccentricity and the orientation of the $\chi^2$ contour plots of the fits, used to derived the uncertainties of the fluxes and slopes. The 68 per cent ellipsoids have an average eccentricity of $0.91\pm0.03$ and they are tilted on average by an angle of $(79\pm2)\degr$ counter-clockwise from the horizontal (flux) axis favouring, if at all, trends moving the opposite direction i.e.\ a positive correlation between flux and $\Gamma$. We therefore conclude that for \n7, contrary to what is observed in other luminous Seyfert galaxies, the X-ray photon index anti-correlates with the X-ray flux.

\subsection{Hardness Ratio Analysis}
\label{ssect:hr}
\begin{figure}
\includegraphics[width=3.5in]{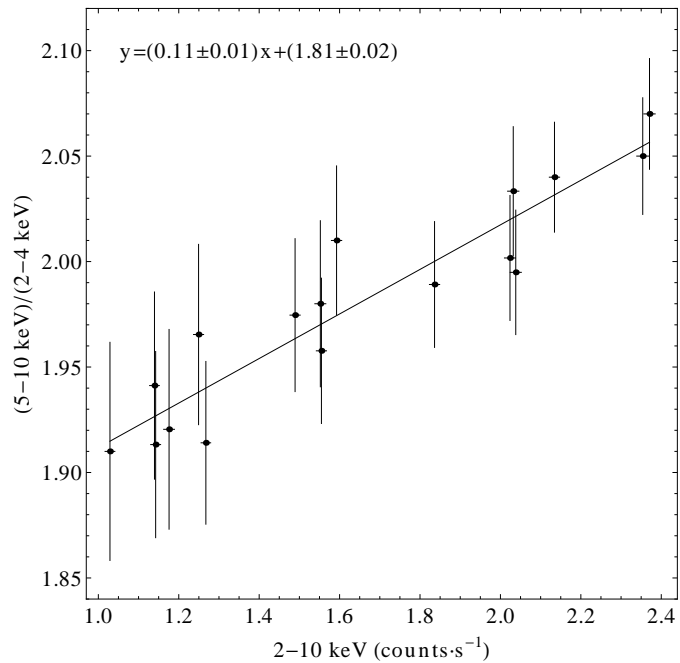}
\caption{Hardness ratio plot of the count-rate in (5--10 keV)$/$(2--4 keV) versus 2--10 keV. The bin size corresponds to 50 consecutive observations. The solid line indicates the best-fit linear model to the data taking into account the uncertainties in both coordinates. The positive slope implies a `harder when brighter' X-ray spectral behaviour.}
\label{fig:rxte_hardnessRatio}
\end{figure}

A completely model independent way to check for the validity of the above-mentioned anti-correlation behaviour, is by estimating the hardness ratio from the X-ray light curves. After binning the light curves in bins of 50 consecutive observations, we estimate the hardness ratio (5--10 keV)$/$(2--4 keV) versus the overall count-rate in 2--10 keV. In Fig.~\ref{fig:rxte_hardnessRatio} we plot the corresponding estimates, showing a clear increasing trend which implies that the X-ray behaviour of the source becomes `harder' when the source gets brighter.\par
We can now fit to the hardness ratio data a linear model, $(y=\alpha x+\beta)$, taking into account the errors in both coordinates (as in Section \ref{ssect:rxte_lc}). The best fit model has a slope of $\alpha=0.11\pm0.01$ and an intercept $\beta=1.81\pm0.02$ yielding a $\chi^2$ of 4.31 for 15 d.o.f. having a null hypothesis probability of $3.4\times10^{-3}$. Since the 99 per cent confidence interval for the slope is (0.07,0.14) the null hypothesis can be rejected at 1 per cent significance level. Finally, the value of ${\itl t}$-statistic that we get from the data is 10.1\footnote{The value of ${\itl t}$-statistic is ${\itl t}_{15,0.005}=2.95$}, corresponding to a probability of getting such a value from chance alone equal to $4.6\times10^{-8}$. Therefore, we can robustly conclude that the best-fit slope is significantly different from zero.

\subsection{The nuclear broad-band Spectral Energy Distribution}
\label{ssect:SED}
\begin{figure}
\includegraphics[width=3.5in]{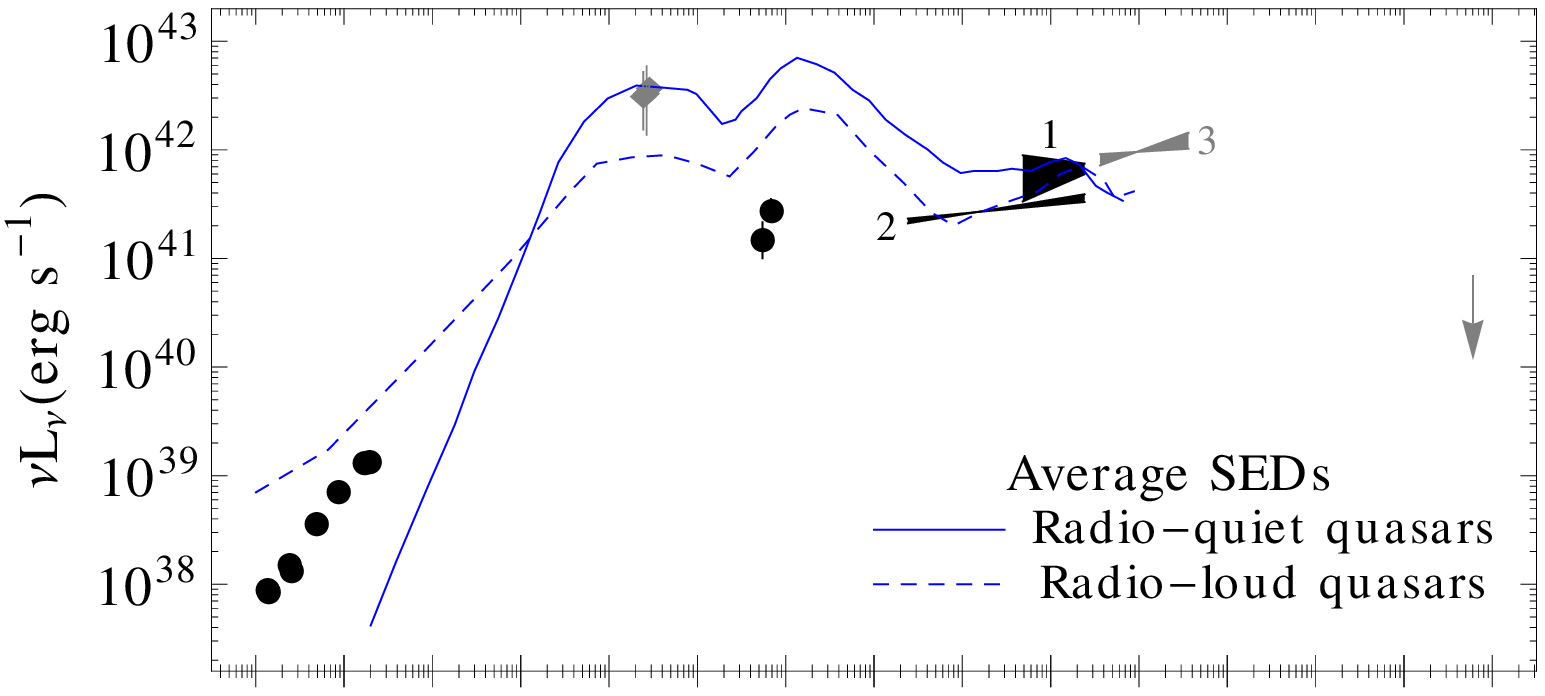}\\[-1.73em]
\hspace*{0em}\includegraphics[width=3.5in]{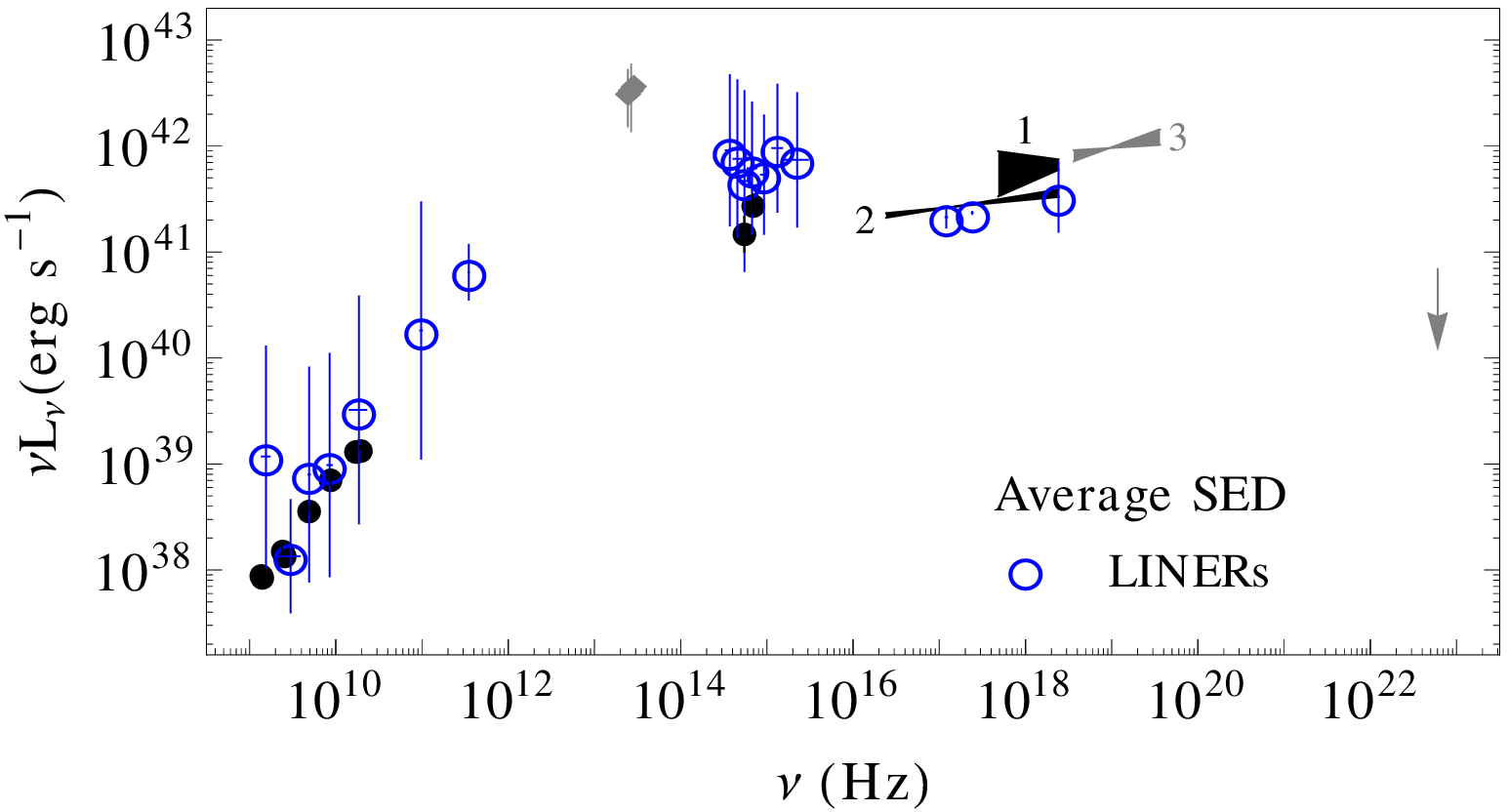}
\caption{The SED of \n7. For both panels, black symbols correspond to flux density estimates derived in this work and grey symbols correspond to archival data. The enumerated symbols, 1,2 and 3, correspond to the average RXTE spectrum (depicting the long-term variations), \textit{XMM-Newton} and \textit{Swift}-BAT spectrum, respectively. The grey arrow, corresponds to the \textit{Fermi}-LAT upper-limit, in the 0.1-100 GeV energy range. (Top panel) The SED of \n7 together with the average SEDs of radio-quiet and radio-loud quasars shown with the blue solid- and dashed-lines, respectively \citep[taken from][]{elvis94}. (Bottom-panel) The SED of \n7 together together with the average SEDs LINERs shown with blue open-circles \citep[taken from][]{eracleous10}.}
\label{fig:SEDngc7213}
\end{figure}

The average spectral energy distribution of \n7\ (in $\nu L_{\nu}$ representation) is shown in Fig.~\ref{fig:SEDngc7213} (both panels) with black symbols corresponding to flux estimates (listed in Table \ref{tab:fluxEntries}) derived in this work and and grey symbols to archival data. The error bars correspond to the standard deviation around the mean flux, at a given frequency, whenever multiple observations from different epochs are available. In this way, the error bars are indicative of the actual variability of \n7 (the actual uncertainty of single pointing observations is much smaller than the symbol size). The average flux from the ensemble of \textit{RXTE} observations in the 2--10 keV energy range is shown by symbol 1, and corresponds to a power law having normalization of $(6.51\pm3.73)\times 10^{-3}$ photons s$^{-1}$ cm$^{-2}$  keV$^{-1}$ (at 1 keV) and a photon index of $1.87\pm0.23$. The bowtie 2 indicates the 0.1--10 keV spectrum from the \textit{XMM-Newton} observation (Emmanoulopoulos et al.\,in prep.) which is broadly consistent with a power law with a normalization $2.99^{+0.06}_{-0.03}\times10^{-3}$ photons s$^{-1}$ cm$^{-2}$ keV$^{-1}$ (at 1 keV) and a photon index $1.88^{+0.04}_{-0.02}$. The bowtie 3 indicates the average flux estimate registered by the \textit{Swift}-BAT instrument, between 15--150 keV, corresponding to a power law with normalization\footnote{The flux value is consistent with the one derived from the maps of the 4\textsuperscript{th} \textit{IBIS/ISGRI} soft $\gamma$-ray survey catalogue \citep{bird10}.} $5.4^{+3.1}_{-2.0}\times 10^{-3}$ photons s$^{-1}$ cm$^{-2}$ keV$^{-1}$ (at 1 keV) and a photon index of 1.82$^{+0.13}_{-0.12}$.  Finally, the 0.1--100 GeV the \textit{Fermi}-LAT upper limit, assuming a photon index of 1.75, is $3\times10^{-9}$ photons s$^{-1}$ cm$^{-2}$, is indicated by the arrow at 0.25 GeV (mean energy of the {Fermi}-LAT energy band).\par
The mean SED of \n7 is shown in Fig.~\ref{fig:SEDngc7213}. This is true even for the {\it RXTE} and {\it Swift-}BAT data, since at these flux levels the contribution of any host galaxy emission should be minimal, especially in the {\it Swift-}BAT band, as well as for the optical data, where we have carefully discard the host galaxy contribution from our measurements. Also note that, the optical, {\it RXTE} and {\it Swift}-Bat data plotted on this figure are indicative of the source flux averaged over many years, which are largely overlapping. On the other hand, the beam size of the radio observations is rather large, due the short exposure times (Section \ref{ssect:sed_data}). As a result, the observed radio fluxes we have used may be contaminated by radio emission from a substantial part of the host galaxy itself. However, our radio flux estimates, at least for the 4.8 and 8.6 GHz bands, are identical to the mean flux values reported by \citet{bell11}, who used a beam size of $0.5\arcsec$ for a 12 h integration time. In addition, the observed variability at all radio bands indicate that most of the emission should originate from the nucleus itself. Detailed X-ray studies \citep[e.g. Emmanoulopoulos et al.\,in prep.,][]{lobban10} indicate that the intrinsic absorption towards the nuclear source in \n7 is minimal in addition to the fact that the galaxy of \n7 is face-on.\par
Consequently, the SED in Fig.~\ref{fig:SEDngc7213} should be representative of the mean-intrinsic nuclear SED of \n7. 
The blue solid and dashed-line in the top panel of Fig.~\ref{fig:SEDngc7213} indicates the average SED of radio-quiet and radio-loud quasars \citep{elvis94}, respectively, and the blue open-circles in the bottom panel of the same figure indicate the average SED of LINERs \citep{eracleous10}. Their relative position is set by minimizing the distance of their logarithmic ordinates from those of \n7's, at the same frequencies, weighted by their squared errors. Since the distance minimization procedure is done in the logarithmic plane, in this way we estimate effectively an optimum normalization for each average-SED. Our results suggest that it is the LINERs mean SED which fits best, within the \n7's SED. We therefore conclude that the nuclear continuum emission of the source is indeed representative of the mean SED of nearby LINER's.\par 
Despite the fact that the (archival) near-IR flux measurements of \n7 (as shown in the bottom panel of Fig.~\ref{fig:SEDngc7213}) are representative of emission from a central region which is less than $0.35\arcsec$ in size, it is not clear whether they correspond to emission form the nuclear source itself. For example, if we indeed observe the active nucleus directly, without any significant absorption, the detected near-IR emission cannot be due to optical/UV nuclear emission being absorbed by the putative dusty, obscuring torus in this galaxy and re-emitted in the near-IR, because the average optical luminosity is significantly lower than the observed near-IR luminosity. On the other hand, \n7 shows evidence of a  starburst driven wind \citep{bianchi08} and hosts a star-forming ring few kpc away from the nucleus. A clumpy torus with toroidal shape \citep{nenkova08}, having a small angular width parameter can then explain the observed infrared luminosity as being due to reprocessed radiation from star-forming regions, while the optical nuclear emission could escape unabsorbed through a torus-hole. Note that LINERs are thought to have relatively hot yet normal main-sequence O stars \citep{shields92} able to heat the dust around them, something that strengthens even more the LINER interpretation of \n7 SED.\par
The SED plotted in Fig.~\ref{fig:SEDngc7213} is based on more observations of the nuclear source flux, and is spread over a larger frequency range, than the SED presented by \citet{starling05}. We can therefore use it to derive a more accurate estimate of the nuclear bolometric luminosity. After interpolating the SED linearly in logarithmic space, we integrate it between 1.344 GHz and $3.63\times10^{19}$ Hz (i.e.\ 150 keV) and we derive $L_{\rm bol}=1.7\times 10^{43}$ ergs s$^{-1}$, yielding an accretion rate of 0.14 per cent of Eddington limit.

\section{Discussion}
\label{sect:discussion}
We have analysed the long-term RXTE observations of \n7 and we found that this low luminosity source exhibits a clear `harder when brighter' X-ray behaviour (Fig.~\ref{fig:rxte_hard_bright}). This is the first time that such a spectral variability behaviour is reported for either a `low' or `high' luminosity AGN, and is contrary to what is observed in luminous Seyferts and quasars. We also constructed the average, nuclear SED of this source, using archival and proprietary data that we reduced, together with other measurements from the literature and we found that its shape closely resembles the mean SED of nearby LINERs. Finally, we provided a new measurement for the nuclear bolometric luminosity of \n7: $L_{\rm bol}=1.7 \times 10^{43}$ ergs s $^{-1}$, yielding a rather low accretion rate of 0.0014 (i.e.\ 0.14 per cent) of the Eddington limit.\par 
The average, broad band nuclear SED of \n7 resembles that of LINERs (Fig.~\ref{fig:SEDngc7213}), bottom panel). In the average SEDs of radio-loud and the radio-quiet AGN, the optical flux is quite higher than the X-ray flux. This is definitely not the case for this source, since the optical flux lies well below that of the X-rays. This is a very robust result, as both the optical and the X-ray data are based on long, monitoring observations, i.e.\ they show the average behaviour of the source over many years. Furthermore, \citet{wu83} noticed that the UV excess for \n7 should be extremely week or absent, while both the recent {\it XMM-Newton} observations (Emmanoulopoulos et al, in prep.), as well as the {\it Suzaku} observations \citep{lobban10} show no indication of soft X-ray excess, which is typical in most luminous AGN. All these results strongly suggest that the `big-blue bump' is missing in this source. This can be expected in a scenario in which the inner part of the geometrically thin and optically thick disc is missing in \n7.\par  
Our results suggest that the accretion rate of the source is significantly smaller (by a factor of 10) than the `critical' rate at which accreting BHXRBs move from the `hard' to the `soft state'. Therefore, \n7 could be the `hard state' analogue of BHXRBs, and in fact, its `harder when brighter' behaviour strongly supports this hypothesis. Although the global relationship between $\Gamma$ and $\xi$ in BHXRBs is well-established by comparing measurements from single-epoch observations \citep[e.g.][]{wu08,younes11} the correlation of $\Gamma$ with $\xi$ on short time scales (within an observation, i.e.\ down to minutes or even seconds) is less well-determined. However, \citet{axelsson08} showed for the BHXRB Cyg\,X-1 that the hardness-flux anti-correlation, seen in variations within an observation in brighter `hard' states, becomes a positive correlation in the faintest `hard' states.  This change in behaviour is consistent with a positive $\Gamma$--$\xi$ correlation at higher $\xi$ changing to an anti-correlation at lower $\xi$, i.e.\ the variations within an observation follow the same trends as the global $\Gamma$--$\xi$ relationship.\par
Theoretically a considerable progress has been accomplished in this field with a better understanding of the complexity of the BHXRBs' `hard state' as well as LLAGN \citep[see for a review,][]{narayan05}. A currently proposed model involves an accretion disc plus a hot accretion flow model, ADAF model, to explain the spectral dependence on accretion rate for BHXRBs, from quiescence up to the `soft state'. For intermediate accretion rates, as the accretion rate increases the Compton parameter in the hot accretion flow increases as well, producing a Comptonisation X-ray spectrum with a `harder' power law slope. This regime is identified with the `hard state', at mass accretion rates of 1--8 per cent of the Eddington rate, since the accretion process there is inefficient corresponding to lower fractions of L$_{\rm Edd}$, consistent with our observations of \n7. As the accretion rate increases further, the model predicts a different behaviour in which increasing luminosity corresponds to `softer' X-ray power law slopes. This regime is identified with the `soft state' and this is the what is normally observed in higher luminosity AGN.\par
Finally, another interesting feature of the source's SED is the fact that its X-ray emission above 20 keV is quite high, and implies a high-energy cut-off larger than 350 keV \citep{lobban10}. \n7 shows some week evidence of a radio jet structure at 8.4 GHz \citep{blank05}. If such a structure is indeed there and it is aligned towards the observers direction it could produce relativistically amplified radio emission through synchrotron radiation, and enhanced X-ray emission through inverse Compton radiation of either the synchrotron photons and/or optical photons from the star-burst environment of the host galaxy. This possibility could explain the fact that the radio emission of the source lies between that of radio-loud and radio-quiet AGN, similar to blazars. At the same time, the existence of a jet could also account for the `harder when brighter' behaviour of \n7, something which is commonly observed in blazars \citep[e.g.][]{krawczynski04,gliozzi06,zhang06} and can be explained in the framework of the synchrotron self Compton models. A jet model has also been proposed to explain the spectral evolution of the BHXRB XTE\,J1550-564 as it moves from the fading state towards the X-ray `hard state' \citep{russell10}.\par
Therefore, a jet pointing towards us, contributing significantly to the radio and X-ray emission of the source can not be ruled out, and in fact, for the case of ADAF models, it can be created through the Blandford-Znajek mechanism \citep{armitage99}. In the `hard state' of BHXRBs the kinetic luminosity in the jet is believed to equal the radiation luminosity \citep{fender10}. For the case of \n7, assuming the existence of a jet, the previously derived accretion rate may therefore be lower by a factor of two, still remaining well below the typical transition between the `hard' and the `soft' state in BHXRBs. In the future, models combining a jet component with a geometrically thin disk and ADAF \citep{yuan05} need to be tested for the case of \n7.

\section*{Acknowledgments}
DE and IMM acknowledge the Science and Technology Facilities Council (STFC) for support under grant ST/G003084/1. IP acknowledges support by the EU
FP7-REGPOT 206469 grant. PA acknowledges support from Fondecyt grant number 11100449. This research has made use of NASA's Astrophysics Data System Bibliographic Services. Finally, we are grateful to the anonymous referee for the useful comments and suggestions that helped improved the quality of the manuscript.

\bsp
\label{lastpage}
\end{document}